\documentstyle[twoside,fleqn,espcrc2]{article}

\font\mybb=msbm10 at 12pt
\def\bb#1{\hbox{\mybb#1}}
\def\bZ {\bb{Z}}

\input epsf.tex

\def\NS{${\rm NS}\otimes{\rm NS}\ $}
\def\RR{${\rm R}\otimes{\rm R}\ $}

\newcommand{\AmS}{{\protect\the\textfont2
  A\kern-.1667em\lower.5ex\hbox{M}\kern-.125emS}}

\title{Brane Surgery}

\author{P. K. Townsend\address{DAMTP, University of Cambridge,\\ 
        Silver St., Cambridge, U.K.}\thanks{To appear in proceedings of the
European Research Conference on {\it Advanvced Quantum Field Theory}, La
Londe les Maures, France, September 1996, held in memory of Claude Itzykson.}}
       
\begin{document}

\begin{abstract}
Some aspects of the role of p-branes in non-perturbative
superstring theory and M-theory are reviewed. It is then shown how the
Chern-Simons terms in D=10 and D=11 supergravity theories determine which 
branes can end on which, i.e. the `brane-boundary rules'.
\end{abstract}

\maketitle

\section{Introduction}

Extended objects, known as `branes', currently play an essential role in
our understanding of the non-perturbative dynamics underlying
ten-dimensional (D=10) superstring theories and the 11-dimensional (D=11)
M-theory (see \cite{schwarz} for a recent review). In the context of the
effective D=10 or D=11 supergravity theory a `p-brane' is a solution of the 
field equations representing a p-dimensional extended source for an abelian
$(p+1)$-form gauge potential $A_{p+1}$ with $(p+2)$-form field strength
$F_{p+2}$. As such, the $p$-brane carries a charge
\begin{equation}
Q_p = \int_{S^{D-p-2}}\!\!\! \star F_{p+2}\ ,
\label{eq:introa}
\end{equation}
where $\star$ is the Hodge dual in the D-dimensional spacetime and the integral
is over a $(D-p-2)$-sphere encircling the brane, as shown schematically in the
figure below:
\vskip 0.5cm
\epsfbox{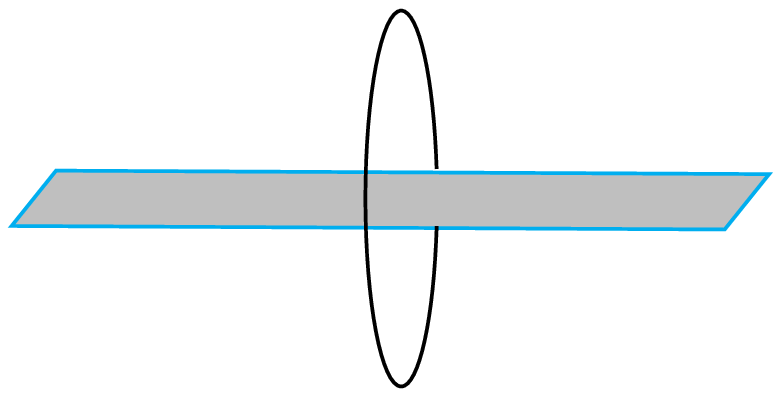} 
\vskip 0.5cm 
\noindent 
In the case of a static infinite planar p-brane this formula is readily
understood as a direct generalization of the $p=0$ case, i.e. a point particle 
in electrodynamics, with the $(D-p-1)$-dimensional `transverse space' (spanned 
by vectors orthogonal to the $(p+1)$-dimensional worldvolume) taking the place 
of space. In the case of a closed p-brane, static or otherwise, the charge $Q_p$
can be understood (after suitable normalization) as the linking number of the
p-brane with the $(D-p-2)$-sphere in the $(D-1)$-dimensional space. 

Examples for $p=1$ are provided by the D=10 heterotic strings, for which
\begin{equation}
Q_1 = \int_{S^7}\! \star H\ ,
\label{eq:introb}
\end{equation}
where $H$ is the 3-form field strength for the 2-form gauge potential $B$ from
the massless Neveu-Schwarz (NS) sector of the string spectrum. Further
D=10 examples are provided by the type II superstrings, with the
difference that $B$ is now the 2-form of \NS origin in type II superstring
theory. A D=11 example is the charge
\begin{equation}
Q_2 = \int_{S^7}\! \star F
\label{eq:introc}
\end{equation}
carried by a supermembrane, where $F=dA$ is the 4-form field strength for the
3-form potential $A$ of D=11 supergravity.

The statement that a $p$-brane carries a charge of the above type can be
rephrased as a statement about interaction terms in the effective
worldvolume action governing the low-energy dynamics of the object. Consider,
for example, type II and heterotic strings. Let
$\sigma^i$ be the worldsheet coordinates and let $X^\mu(\sigma)$ describe the
immersion of the worldsheet in the D=10 spacetime. Then the worldsheet
action in a background with a non-vanishing 2-form $B$ will include the 
term 
\begin{equation}
I_B= \int\! d^2\sigma\; \varepsilon^{ij} \partial_i X^\mu \partial_j X^\nu\
B_{\mu\nu}\big(X(\sigma)\big)\ ,
\label{eq:introd}
\end{equation}
where $\varepsilon^{ij}$ is the alternating tensor density on the worldsheet. 
Thus the string is a source for $B$ and, since the coupling is `minimal', it will
contribute to the charge $Q_1$ defined above. Similarly, in a D=11 background
with non-vanishing 3-form potential $A$, the membrane action includes the
term \cite{bst}
\begin{equation}
I_A= \int\! d^3\xi\; \varepsilon^{ijk} \partial_i X^M \partial_j X^N 
\partial_k X^P A_{MNP}\ ,
\label{eq:introe}
\end{equation}
where $X^M(\xi)$ describes the immersion of the supermembrane's worldvolume in
the D=11 spacetime, and $\xi^i$ are the worldvolume coordinates. This
minimal interaction implies that the membrane is a source for $A$ with
non-vanishing charge $Q_2$. 

The actions $I_A$ and $I_B$ are actually related by double-dimensional reduction,
as are the full supermembrane and IIA superstring actions \cite{DHIS}. The
dimensional reduction to D=10 involves setting $X^M=(X^\mu, y)$ where $y$ is 
the coordinate of the compact 11th dimension, and taking all fields to be
independent of $y$. From the worldvolume perspective this amounts to a special
choice of background for which $k=\partial/\partial y$ is a Killing vector
field. Double-dimensional reduction is then achieved by setting $\xi=
(\sigma,\rho)$ where $\rho$ is the coordinate of a compact direction of the
membrane, and then setting $\partial_\rho X^\mu=0$ and $dy=d\rho$, which is the
ansatz appropriate to a membrane that wraps around the 11th dimension. The
action $I_A$ then becomes $I_B$ after the identification $B={\it i}_kA$, where ${\it i}_k$ indicates contraction with the vector field $k$. 

A coupling to $B$ of the form (\ref{eq:introd}) is possible only for oriented 
strings. Of the five D=10 superstring theories all are theories of oriented
strings except the type I theory. Thus, the type I string does {\it not} couple
minimally to $B$. Instead, it couples {\it non-minimally}. In the
Lorentz-covariant GS formalism in which the worldsheet fermions,
$\theta$, are in a spinor representation of the D=10 Lorentz group, the
worldsheet interaction Lagrangian is
\begin{equation}
L_B = \bar \theta \Gamma^{\mu\nu\rho} \theta H_{\mu\nu\rho}\ .
\label{eq:introf}
\end{equation}
Because of the `derivative' coupling of the string to $B$ through its field
strength $H$, the $Q_1$ charge carried by the type I string vanishes. As the
above interaction shows, the type I string theory origin of $B$ is in the \RR
sector rather than the \NS sector. This example illustrates a general feature of
string theory: \RR charges are {\it not} carried by the fundamental string.
If there is anything that carries the charge $Q_1$ in type I string theory it
must be non-perturbative. It is now known that there is such a
non-perturbative object in type I string theory \cite{dab,hull,polwit}; it
is just the $SO(32)$ heterotic string! This is one of the key pieces of evidence 
in favour of the proposed `duality', i.e. non-perturbative equivalence, of the 
type I and $SO(32)$ heterotic string theories. Another is the fact that the two
effective supergravity theories are equivalent, being related to each other by a
field redefinition that takes $\phi\rightarrow -\phi$, where
$\phi$ is the dilaton \cite{wita,ark}. Since the vacuum expectation value
$\langle e^\phi\rangle$ is the string coupling constant $g_s$ this means that
the weak coupling limit of one theory is the strong coupling limit of the other.

An important consequence of the charge $Q_p$ carried by a $p$-brane is that it
leads to a BPS-type bound on the p-volume tension, $T_p$ of the form
$T_p\ge c_p|Q_p|$, where $c_p$ is some constant characteristic of the particular
supergravity theory, the choice of vacuum solution of this theory, and the value
of $p$. If one considers the class of static solutions with $p$-fold
translational symmetry then a bound of the above form follows from the
requirement that there be no naked singularities. This bound is saturated by the
solution that is `extreme' in the sense of General Relativity, i.e. for which
the event horizon is a degenerate Killing horizon\footnote{When a dilaton is
present this is true only for an appropriate definition of the metric.}.
However, these considerations are clearly insufficient to show that the p-brane
tension actually {\it is} bounded in this way because the physically relevant
class of solutions is the much larger one for which only an appropriate {\it
asymptotic} behaviour is imposed. Remarkably, the attempt to establish a 
BPS-type bound succeeds if and only if the theory is either a supergravity
theory, or a consistent truncation of one \cite{ght}\footnote{In contrast, the
proof of positivity of the ADM mass of asymptotically-flat spacetimes is not
subject to this restriction since, for example, it is valid for arbitrary
D.}. In particular, {\it the presence of various Chern-Simons terms in the
Lagrangians of D=10 and D=11 supergravity theories is crucial to the existence of
a BPS-type bound on the tensions of the $p$-brane solutions of these theories}.
This is so even when, as is usually the case, these Chern-Simons (CS) terms play
no role in the $p$-brane solutions themselves in the sense that they are equally
solutions of the (non-supersymmetric) truncated theory in which the CS terms are
omitted. These facts hint at a more important role for the supergravity CS terms
in determining the properties of $p$-branes than has hitherto been appreciated.
This observation provided the principal motivation for this article, as will
become clear.  

Although the charge $Q_p$ has only a magnitude, it is associated with an object
whose spatial orientation is determined by a $p$-form of fixed magnitude. Thus, 
a $p$-brane is naturally associated with a $p$-form charge of magnitude $Q_p$.
Indeed, the supersymmetrization of terms of the form (\ref{eq:introd}) or
(\ref{eq:introe}) leads to a  type of super-Wess-Zumino term that implies a
modification of the standard supersymmetry algebra to one of the (schematic)
form \cite{azc,pktc}
\begin{equation}
\{Q,Q\} = \Gamma\cdot P + \Gamma^{(p)}\cdot Z_p \ ,
\label{eq:extraone}
\end{equation}
where $\Gamma^{(p)}$ is an antisymmetrized product of $p$ Dirac matrices and
$Z_p$ is a $p$-form charge whose magnitude is given by the coefficient of the
Wess-Zumino term. For $p=0$ this is the well-known modification that
includes $Z_0=Q_0$ as a central charge. More generally, $Q_p$ may be identified
as the magnitude of $Z_p$, and an extension of the arguments used in the $p=0$
case \cite{witol,gh} shows that the supersymmetry algebra (\ref{eq:extraone})
implies the BPS-type bound on the p-brane tension $T_p$. It also shows that the
`extreme' $p$-brane solutions of supergravity theories which saturate the bound
must preserve some of the supersymmetry, and the fraction preserved is always
1/2 for $p$-brane solutions in D=10 and D=11\footnote{There are other solutions
which preserve less than half the supersymmetry, and which have an
interpretation as $p$-branes in $D<10$, but these can always be viewed as
composites (e.g. intersections) of $p$-branes in D=10 or D=11. We shall not need
to consider such solutions here.}. The heterotic and type II superstrings and
the D=11 supermembrane are examples not only of charged $p$-branes but also of
{\it extreme} charged $p$-branes. This follows from the `$\kappa$-symmetry' of
their  Lorentz covariant and spacetime supersymmetric worldsheet/worldvolume
actions (see \cite{pktb} for a review). The BPS-saturated $p$-branes are
important in the context of the non-perturbative dynamics of superstring
theories or M-theory for essentially the same reasons that BPS-saturated
solitons are important in D=4 field theories. In fact, most of the the latter
can be understood as originating in D=10 or D=11 BPS-saturated $p$-branes. For
these reasons, the BPS-saturated p-branes are the ones of most interest and will
be the only ones  considered here. It should therefore be understood in what
follows that by `brane' we mean `BPS-saturated brane'.

One of the lessons of recent years has been that much can be learned about the
non-perturbative dynamics of superstring theories from the effective
D=10 supergravity theories. One example of this is the fact that there exist
p-brane solutions of type II supergravity theories which are charged, in the
sense explained above, with respect to the $(p+1)$-form gauge fields from the \RR
sector of the corresponding string theory. By supposing these \RR branes to be
present in the non-perturbative string theory one can understand how otherwise
distinct superstring theories might be dual versions of the same underlying
theory. The basic idea is that branes can `improve' string theory in
the same way that strings `improve' Kaluza-Klein (KK) theory. For example, the
$S^1$-compactified IIA and IIB supergravity theories have an identical massless
D=9 spectrum but are different as Kaluza-Klein theories because
their massive modes differ. The corresponding string theories are the same,
however, because the inclusion of the string winding
modes restores the equivalence of the massive spectra. Similarly, the
$K_3$-compactified type IIA superstring theory and the $T^4$-compactified
heterotic string theory have an identical massless D=6 spectrum, but since
they differ in their perturbative massive spectra they are inequivalent as
perturbative string theories. However, the non-perturbative massive spectrum of
the IIA superstring includes `wrapping' modes of 2-branes around
2-cycles of $K_3$ \cite{townhull}. The inclusion of these leads to the same
massive BPS spectrum in the two theories, and there is now strong evidence of
a complete equivalence \cite{wita,vafwit,vafa,sen}. This evidence rests, in
part, on  the fact that \RR branes now have a remarkably simple description
\cite{pol} in string theory as D-branes, or D-$p$-branes if we wish to specify
the value of
$p$; the worldvolume of a D-$p$-brane is simply a (p+1)-dimensional hyperplane
defined by imposing $(D-p-1)$ Dirichlet boundary conditions at the boundaries of
open string worldsheets.

The distinctions between the various kinds of type II p-brane, such as whether
they are of \NS or \RR type, are not intrinsic but are rather artefacts (albeit
very useful ones) of perturbation theory. Non-perturbatively, all are on an 
equal footing since any one can be found from any other one by a combination of
`dualities'. This feature is apparent in the IIA or IIB effective supergravity
theories which treat all p-form gauge fields `democratically'. Their string
theory origin is nevertheless apparent from the supergravity solutions if the
latter are expressed in terms of the {\it string metric}, instead of the
canonical, or `Einstein', metric. One then finds that there are three categories
of p-brane, `fundamental' (F), `Dirichlet' (D) and `solitonic' (S) according to
the dependence of the p-volume tension on the string coupling constant 
$g_s\equiv\langle e^\phi\rangle$. Specifically,
\begin{equation}
T\sim \cases{1 & {\rm for a fundamental string}\cr
1/g_s & {\rm for a Dirichlet p-brane}\cr
1/g_s^2 & {\rm for a solitonic 5-brane}\ . }
\label{eq:introg}
\end{equation}
Note that according to this classification only strings can be `fundamental'.
This is hardly surprising in view of the fact that we are discussing the
dependence of the tension in terms of the string metric, but it seems to be a
reflection of a more general observation \cite{wita,hullb} that a sensible
perturbation theory can be found only for particles or strings. Similarly only
5-branes can be `solitonic'. This is a reflection of the electric/magnetic
duality between strings and 5-branes in D=10 and the fact that the magnetic dual
of a fundamental object is a solitonic one. It does not follow from this that all
strings are fundamental and all 5-branes solitonic. This is nicely illustrated
by the string solution of N=1 supergravity representing the heterotic string.
This solution is `fundamental' as a solution of the effective supergravity
theory of the heterotic string, as it must be of course, but it is a D-string
when viewed as a solution of the effective supergravity theory of the type I
string. Thus, a single supergravity solution can have two quite different string
theory interpretations. 

The M-theory branes, or `M-branes', consist of only the D=11 membrane and its
magnetic dual, a fivebrane. We saw earlier that the classical IIA superstring
action is related to that of the D=11 supermembrane by double-dimensional
reduction. This was widely considered to be merely a `coincidence', somewhat
analogous to the fact that IIA supergravity happens to be the dimensional
reduction of D=11 supergravity; after all, the {\it quantum} superstring theory
has D=10 as its critical dimension. However, the critical dimension
emerges from a calculation in {\it perturbative} string theory. It is still
possible that the non-perturbative theory really is 11-dimensional, but if this
is so the KK spectrum of the $S^1$-compactified D=11 supergravity must appear in
the non-perturbative IIA superstring spectrum. It was pointed out in
\cite{tow,wita} that the extreme black holes of IIA supergravity, now regarded
as the effective field theory realization of D-0-branes, are candidates for this
non-perturbative KK spectrum. This means that the IIA superstring really {\it
is} an $S^1$-wrapped D=11 supermembrane, but it does not then follow that the
supermembrane is also `fundamental' because this adjective is meaningful only in
the context of a specific perturbation theory. For example, the $SO(32)$
heterotic string is `fundamental' at weak coupling but as the coupling increases
it transmutes into the D-string of the type I theory. Another example is the IIB
string which is `fundamental' at weak coupling but which transmutes into the
D-string of a dual IIB theory at strong coupling \cite{jhs,wita}. In the IIA
case the strong coupling limit is a decompactification limit in which the D=11
Lorentz invariance is restored and the effective D=10 IIA supergravity is
replaced by D=11 supergravity \cite{wita}. The `fundamental' IIA superstring
transmutes, in this limit, into the unwrapped D=11 membrane of M-theory but,
because of the absence of a dilaton, there is no analogue of string
perturbation theory in D=11 and so there is no analogous basis for deciding 
whether or not the membrane is `fundamental'. Nevertheless, as we shall shortly
see, there is an intrinsic asymmetry between M-theory membranes and fivebranes
which suggests a fundamental role for the membrane in some as yet unknown
sense\footnote{This is also suggested by the `n=2 heterotic string' approach to
M-theory \cite{KM}.}. 

Given that the heterotic string appears as a D-brane in type I string theory one
might wonder whether the type I string should make an appearance somewhere in
the non-perturbative SO(32) heterotic string theory. As we have seen, however,
the type I string carries no $Q_1$ charge, so its description in the effective
supergravity theory would have to be as a non-extreme, or `black', string.
Infinite uncharged black strings have been shown to be
unstable against perturbations that have the tendency to break the string into
small segments \cite{greg} (whereas extreme strings are stable because they
saturate a BPS-type bound). This is exactly what one expects from string theory
since a closed type I string can break, i.e. type I string theory is a theory of
both closed and open strings. The reason that this is possible for type I
strings, but not for heterotic or type II strings, is precisely that the type I
string carries no $Q_1$ charge. To see this, suppose that a string carrying a
non-zero $Q_1$ charge were to have an endpoint. One could then `slide off' the
7-sphere encircling the string and contract it to a point. Provided that the
integral defining $Q_1$ is homotopy invariant, which it will be if $d\star H=0$,
the charge $Q_1$ must then vanish, in contradiction to the initial assumption. We
conclude that the only breakable strings are those for which $Q_1=0$. Thus type
II and heterotic strings cannot break. Clearly, similar arguments applied to
$p$-branes carrying non-zero $Q_p$ charge lead to the conclusion that they too
cannot break.

By `break' we mean to imply that the $(p-1)$-brane boundary created in this
process is `free' in the sense that its dynamics is determined entirely by the
$p$-brane of which it is the boundary. An `unbreakable' p-brane may nevertheless
be open if its boundary is tethered to some other object because there may then
be an obstruction to sliding the $(D-p-2)$-sphere off the end of the p-brane. 
Examples of such obstructions are the D-branes on which type II superstrings can
end. One way to understand how this is consistent with
conservation of the charge $Q_1$ is to consider the D-brane's effective
worldvolume action, which governs its low-energy dynamics. The field content of
this action is found from the massless sector of an open type II superstring
with mixed  Neumann/Dirichlet boundary conditions at the ends. These fields are
essentially the same as those of the open type I string without Chan-Paton
factors with the difference that they depend only on the D-brane's worldvolume
coordinates (see e.g. \cite{polb}). In particular, these worldvolume fields
include an abelian 1-form potential $V$. The bosonic sector of the effective
worldvolume action, in a general \NS background, can be deduced from the
requirement of conformal invariance of the type II string action for a
worldsheet with a boundary \cite{leigh}. This effective worldvolume action is
found to contain the term
\begin{equation}
-{1\over4} \int d^{p+1}\xi\; |dV-B|^2\ ,
\label{eq:introh}
\end{equation}
where the integral is over the $(p+1)$-dimensional worldvolume $W$ and it is to
be understood that the spacetime 2-form $B$ is pulled back to $W$. 
This shows that the D-brane is a source of $B$. If we modify the equation
$d\star H=0$ in order to include this source we find, by integration, that 
\begin{equation}
Q_1 = \int_{S^{p-2}}\!\! * dV\ ,
\label{eq:introi}
\end{equation}
where $Q_1$ is defined as before in (\ref{eq:introa}). The integral on the right
hand side of (\ref{eq:introi}) is over a $(p-2)$-sphere in the D-brane
surrounding the string's endpoint and $*$ is the {\it worldvolume} Hodge dual.
This result can be interpreted as the statement that the charge of the string
can be `transferred' to an electric charge of a particle on the D-brane, so
charge conservation is compatible with the existence of an open string provided
that its endpoints are identified with charged particles living on a D-brane.

A similar analysis can be applied to the D=11 membrane which, we recall, is an
electric-type source for the 3-form gauge potential $A$ of D=11 supergravity.
In this case, the D=11 fivebrane has a worldvolume action containing the terms
\cite{pkt,aha}
\begin{equation}
-{1\over12}\int d^6\xi\big\{ |{\cal F}_3|^2 -\varepsilon^{ijklmn}
A_{ijk}\partial_l V_{mn}\big\}\ ,
\label{eq:introj}
\end{equation}
where ${\cal F}_3= (dV_2 -A)$ is the 3-form field strength for a worldvolume
2-form potential $V_2$, and it is again to be understood that $A$ is the
pullback of the spacetime field to the worldvolume. Actually, the
worldvolume 3-form 
${\cal F}_3$ is {\it self-dual}, but this condition must be imposed after
variation of the action (\ref{eq:introj}). The second term in this action is
needed for consistency of the self-duality condition with the $V_2$ field
equation. Apart from this subtlety, we see from its worldvolume action that the
fivebrane is a source for
$A$. Its inclusion in the field equation for $A$ leads, after integration, to the
equation
\begin{equation}
Q_2 = \int_{S^3}\! * dV_2\ ,
\label{eq:introl}
\end{equation}
which can be interpreted as the statement that the membrane charge can be
transferred to a charge carried by a self-dual string within the fivebrane.
This string is just the boundary of an open membrane. Thus, the D=11 fivebrane is
the M-theory equivalent of a D-brane \cite{pkt,strom}. 

The above analysis can be generalized \cite{strom} to determine whether a
$p$-brane can end on a $q$-brane, as follows. One first determines the
worldvolume field content of the $q$-brane. If this includes a $p$-form gauge
field $V_p$, and if the spacetime fields include a $(p+1)$-form gauge potential
$A_{p+1}$, then one can postulate a coupling of the form $\int |dV_p
-A_{p+1}|^2$ in the $q$-brane's effective worldvolume action. This leads to the
$q$-brane appearing as a source for $A_{p+1}$ such that 
\begin{equation}
Q_p= \int_{S^{q-p}}\!\! * dV_p\ ,
\label{eq:introm}
\end{equation}
where the integral in the $q$-brane is over a $(q-p)$-sphere surrounding the
$(p-1)$-brane boundary of the p-brane. Thus, the $p$-brane charge can be
transferred to the electric charge of the $(p-1)$-brane boundary living in the
$q$-brane. That is, charge conservation now permits the $p$-brane to be open
provided its boundary lies in a $q$-brane. The cases discussed above clearly fit
this pattern, but there are drawbacks to this approach. Firstly, it is indirect
because one must first determine the worldvolume field content of all relevant
branes. Secondly, it is {\sl ad hoc} because, in general, the worldvolume
coupling is postulated rather than derived. The subtleties alluded to above in
the construction of the fivebrane action show that this is not a trivial matter.
In fact, even the bosonic fivebrane action is not yet fully known and until it
is one cannot be completely certain that the wanted terms in this action
really are present. 

In this contribution I will present a new, and extremely simple, method
for the determination of when $p$-branes may have boundaries on $q$-branes.
Essentially, {\it one can read off from the Chern-Simons terms in the
supergravity action whether any given $p$-brane can have a boundary and, if so,
in what $q$-brane the boundary must lie}. As such, the method provides a further
example of how much can be learned about the non-perturbative dynamics of
superstring theories, or M-theory, from nothing more than the 
effective supergravity theory. I have called the method `brane surgery' because
of a notional similarity to the way in which manifolds can be `glued' together
by the mathematical procedure known as `surgery', but it is not intended that
the term should be understood here in its technical sense. 

It is pleasure to
dedicate this contribution to the memory of Claude Itzykson, who would surely
have apreciated the remarkable confluence of ideas that has marked 
recent advances in the theory that is still, misleadingly, called `string
theory'.


\section{IIB brane boundaries}

I shall explain the `brane surgery' method initially in the context of the IIB
theory. Both IIA and IIB supergravity have in common the bosonic fields
($g_{\mu\nu}, \phi, B_{\mu\nu}$) from the \NS sector, all of which have already
made an appearance above. The remaining bosonic fields come from the \RR sector.
The (massless) \RR fields of the IIB theory are
\begin{equation}
(\ell, B'_{\mu\nu}, C^+_{\mu\nu\rho\sigma})\ ,
\label{eq:onea}
\end{equation}
i.e. a pseudoscalar $\ell$, another 2-form gauge potential $B'$ and a 4-form
gauge potential $C^+$ with a {\it self-dual} 5-form field strength $D^+$. The
self-duality condition makes the construction of an action problematic but, as
with the self-duality condition on the D=11 fivebrane's worldvolume field
strength ${\cal F}_3$, one can choose to impose this condition {\it after}
varying the action. When the IIB action is understood in this way it contains
the CS term\footnote{The conventions can be chosen such that the coefficient is
as given.}
\begin{equation}
C^+ \wedge H\wedge H'\ ,
\label{eq:oneb}
\end{equation}
where $H=dB$, as before, and $H'=dB'$. This CS term modifies the $B$, $B'$ and
$C^+$ field equations. 

Consider first the $B$ equation. This becomes
\begin{equation}
d\star H =  - D^+ \wedge H'\ ,
\label{eq:onec}
\end{equation}
where $D^+=dC^+$ is the self-dual 5-form field strength for $C^+$. 
This can be rewritten as
\begin{equation}
d(\star H - D^+ \wedge B') =0 \ .
\label{eq:oned}
\end{equation}
Since $\star H$ is no longer a closed form its integral over a 7-sphere will no
longer be homotopy invariant. Clearly, the well-defined, homotopy invariant,
charge associated with the fundamental IIB string is {\it not} $Q_1$ as defined
in (\ref{eq:introa}) but rather
\begin{equation}
\hat Q_1 = \int_{S^7}[\star H - D^+ \wedge B']\ .
\label{eq:onee}
\end{equation}
Let us again suppose that the IIB string has an endpoint. Far away from this
endpoint we can ignore all fields other than $H$, to a good approximation, so
that $\hat Q_1\approx Q_1$. How good this aproximation is actually depends on
the ratio of the radius $R$ of the 7-sphere to the its distance $L$ from the
end of the string, and it can be made arbitrarily good by increasing $L$ for
fixed $R$. This shows, in particular, that $\hat Q_1\ne0$ for the fundamental 
IIB string. Let us now `slide' the 7-sphere along the string towards the
endpoint. If the $D^+\wedge B'$ term could be entirely ignored we would be back
in the situation described previously in which we arrived at a contradiction,
so we are forced to suppose that an endpoint is associated with a non-vanishing
value of $D^+ \wedge B'$. Nevertheless, the approximate equality of $\hat Q_1$
to $Q_1$, which ignores the $D^+ \wedge B'$ term, can be maintained, to the
same precision, if the 7-sphere surrounding the string is contracted as we
approach the endpoint so as to keep the ratio $R/L$ constant. Then, as
$L\rightarrow 0$ so also $R\rightarrow 0$, until at $L=0$ the 7-sphere is
contracted to the endpoint itself. We can then deform the 7-sphere into the
product $S^5\times S^2$ so that $\hat Q_1$ now receives its entire contribution
from the $D^+ \wedge B'$ term as follows:
\begin{equation}
\hat Q_1 = -\int_{S^5}D^+\ \times\,\int_{S^2} B'\ .
\label{eq:onef}
\end{equation}
The $S^5\times S^2$ integration region is illustrated schematically by the 
figure below:
\vskip 0.5cm
\epsfbox{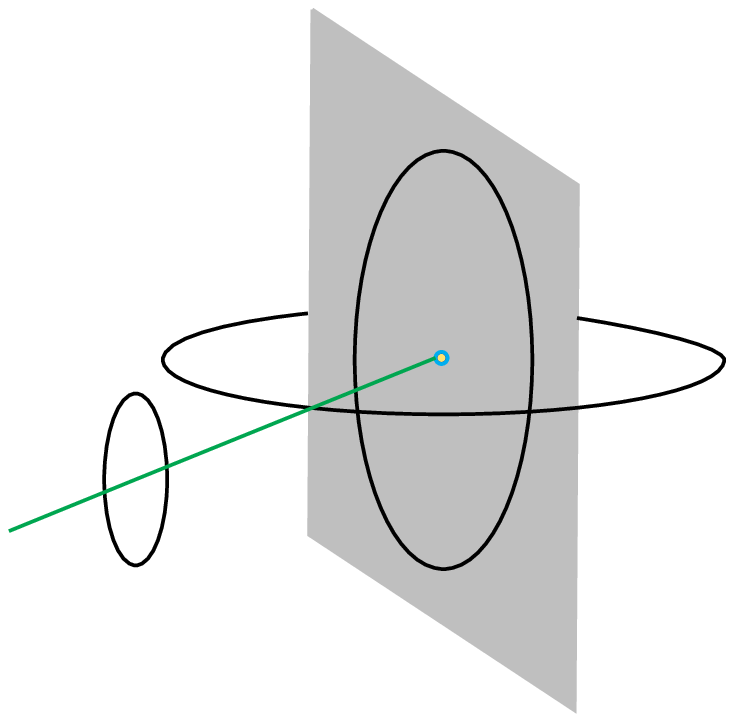} 
\vskip 0.5cm 
\noindent 
Observe that the $S^5$ integral is just the definition of the charge $Q_3$
carried by a 3-brane, so the IIB string has its endpoint on a 3-brane; the
$S^2$ integration surface lies within the 3-brane and surrounds the string
endpoint. Let us choose $Q_3=1$. If we further suppose that $H'\equiv dB' =0$
within the 3-brane, which is reasonable in the absence of any D-string source
for this field, then $B'$ is a closed 2-form which we may write, locally, as
$B'=dV'$ for some 1-form $V'$. Then
\begin{equation}
\hat Q_1 = -\int_{S^2}\! dV'\ .
\label{eq:onefa}
\end{equation}
Effectively, $V'$ is a field living on the worldvolume of the 3-brane. 
Clearly, it cannot be globally defined because the right hand side of 
(\ref{eq:onefa}) is 
a magnetic charge on the 3-brane associated with the vector potential $V'$. 

Now consider the $B'$ equation. Taking the CS term (\ref{eq:oneb}) into account
we have
\begin{equation}
d\star H = D^+ \wedge H\ .
\label{eq:onefb}
\end{equation}
By the same reasoning as before we deduce that the D-string can end on a
3-brane. Charge conservation is satisfied because the D-string charge can be
expressed as
\begin{equation}
\hat Q'_1 = Q_3 \times \int_{S^2}\! B\ .
\label{eq:onefc}
\end{equation}
Since there is no fundamental string source in the problem we may suppose that
$H=0$, so that now $B$ is a closed 2-form which we may write, locally, as
$B=dV$.  For $Q_3=1$ we now have
\begin{equation}
\hat Q'_1 = \int_{S^2}\! dV\ ,
\label{eq:oneg}
\end{equation}
so the D-string charge has been transferred to a magnetic charge of the 1-form
potential $V$ on the 3-brane's worldvolume. 

It must be regarded as a weakness of the above analysis that it does not
supply the relation between $V$ and $V'$, although we know that there must be
one because both supersymmetry and an analysis of the small fluctuations about
the 3-brane solution show that there is only {\it one} worldvolume 1-form
potential. In fact, $V$ and $V'$ are dual in the sense that
\begin{equation}
dV' = *dV\ ,
\label{eq:onega}
\end{equation}
where we recall that $*$ indicates the {\it worldvolume} Hodge dual. 
Using this relation, (\ref{eq:oneg}) becomes
\begin{equation}
\hat Q_1 = \int_{S^2}* dV\ ;
\label{eq:onegb}
\end{equation}
i.e. the endpoint of the IIB string on the 3-brane is an {\it electric}
charge associated with $V$. We thereby recover the D-brane picture for the IIB
3-brane; the fact that the D-string can end on the magnetic charge associated
with $V$ is then a consequence of the strong/weak coupling duality in IIB
superstring theory interchanging the fundamental string with the D-string.
It will be seen from the examples to follow that the need to impose a
condition of the type (\ref{eq:onega}) is a general feature, which is not
explained by  the `brane surgery' method. However, the method does determine
whether a given p-brane can have a boundary and, if so, the possible q-branes in
which the boundary must lie. 

As a further illustration we now observe that whereas (\ref{eq:onec}) was
previously rewritten as (\ref{eq:oned}), we could instead rewrite it as
\begin{equation}
d(\star H + H'\wedge C^+)=0\ .
\label{eq:onegc}
\end{equation}
Thus an equivalent definition of $\hat Q_1$ is
\begin{equation}
\hat Q_1 = \int_{S^7} [\star H +  H'\wedge C^+]\ .
\label{eq:onegd}
\end{equation}
Proceeding as before, but now deforming the $S^7$ into the product $S^3\times
S^4$, we can express $\hat Q_1$ as
\begin{equation}
\hat Q_1 = \int_{S^3} H' \times \int_{S^4} C^+\ .
\label{eq:oneh}
\end{equation}
We recognise the first integral as the D-5-brane charge $Q'_5$. Setting
$Q'_5=1$ and $D^+=0$, we conclude that
\begin{equation}
\hat Q_1 =  \int_{S^4} dV_3\ ,
\label{eq:oneha}
\end{equation}
where $V_3$ is a locally-defined 3-form field on the 5-brane worldvolume, which
can be traded for a 1-form potential $V$ via the relation 
\begin{equation}
dV_3 = * dV\ . 
\label{eq:onehb}
\end{equation}
We conclude that the CS term allows the fundamental IIB string to end on a
5-brane as well as on a 3-brane, and that the end of the string is
electrically charged with respect to a 1-form potential $V$ living on the
5-brane's worldvolume. This is just the usual picture of the D-5-brane.
Interchanging the roles of $B$ and $B'$ leads to the further possibility of the
D-string ending on the solitonic 5-brane.

We have not yet exhausted the implications of the CS term (\ref{eq:oneb}) because
we have still to consider how it affects the $C^+$ equation of motion. We find 
that\footnote{The same equation follows, given the self-duality of $D^+$, from
the `modified' Bianchi identity for $D^+$.}
\begin{equation}
d\star D^+ = -H \wedge H'
\label{eq:onei}
\end{equation}
or
\begin{equation}
d(\star D^+  + H' \wedge B) =0\ .
\label{eq:onej}
\end{equation}
This means that the 3-brane charge should be modified to
\begin{equation}
\hat Q_3 = \int_{S^5} [ \star D^+ + H' \wedge B]\ .
\label{eq:onek}
\end{equation}
This reduces to the previously-defined 3-brane charge $Q_3$ if the 5-sphere
surrounds a 3-brane sufficiently far from the boundary. As before the 5-sphere
can be slid towards, and contracted onto, the boundary, after which it emerges
as the product $S^3\times S^2$. Setting $B=dV$ again we arrive at the expression
\begin{equation}
\hat Q_3 = \int_{S^3}H' \times \int_{S^2} dV
\label{eq:onel}
\end{equation}
for the 3-brane charge. The singularity involved in this deformation of the
7-sphere is now the 2-brane boundary of the 3-brane within a D-5-brane, since we
recognise the first integral on the right hand side of (\ref{eq:onel}) as $Q'_5$.
Setting
$Q'_5=1$ we learn that the 3-brane charge can be transferred to a magnetic
charge of a D-5-brane worldvolume 1-form potential $V$, defined by a 2-sphere in
the D-5-brane surrounding the 2-brane boundary. The main point in all this is
that a 3-brane can have a boundary in a D-5-brane, as pointed out in
\cite{strom}. In fact, this possibility follows by T-duality from the previous
results: the configuration of a D-string ending on a D-3-brane is mapped to a
D-3-brane ending on a D-5-brane by T-duality in two directions orthogonal to
both the D-string and the D-3-brane. By interchanging the roles of $B$ and $B'$
in the above analysis one sees that a 3-brane can also end on a solitonic
5-brane.
 
We have seen that the CS term (\ref{eq:oneb}) allows a IIB string to end on a
D-3-brane or a D-5-brane, but we know from string theory that it can also end on
a D-string or a D-7-brane. As we shall see shortly, these possibilities are
consequences of the fact that the kinetic term for $H'$ actually has the form
\begin{equation}
-{1\over6}|H'- \ell H|^2\ .
\label{eq:onem}
\end{equation}
There is no obvious relation to CS terms yet, but if we perform a duality
transformation to replace the 2-form $B'$ by its 6-form dual $\tilde B'$ with
7-form field strength $\tilde H'$, so that {\it on shell} 
\begin{equation}
\tilde H' =\star H'\ ,
\label{eq:onema}
\end{equation}
then one finds that the dualized action contains the CS term
\begin{equation}
\ell \tilde H' \wedge H\ .
\label{eq:onen}
\end{equation}
Clearly, this modifies the $B$ equation so that, following the steps explained
previously, we end up with an expression
\begin{equation}
\hat Q_1 = \int_{S^7}\tilde H' \times\int_{S^{0}} \ell\ .
\label{eq:oneo}
\end{equation}
The first integral can be identified, using (\ref{eq:onema}), as the D-string
charge. The final `integral' over $S^0\equiv Z_2$ is just the difference between
the value of $\ell$ on either side of the string boundary on the D-string; by the
same logic as before we may assume that $d\ell=0$, {\it locally}, but allow the
constant $\ell$ to be different on either side. Thus, the charge $Q_1$ on the
fundamental IIB string is transformed into the topological charge of a type of
`kink' on the D-string. 

Alternatively, we can deform $S^7$ to $S^1\times S^6$, so that
\begin{equation}
\hat Q_1 = -\int_{S^1} d\ell \times \int_{S^{6}} \tilde B'\ .
\label{eq:onep}
\end{equation}
The first integral is the charge $Q_7$ associated with the D-7-brane. This
charge can be non-zero because of the periodic identification of $\ell$
implied by the conjectured $Sl(2;\bZ)$ invariance of IIB superstring theory
\cite{townhull}.  For $Q_7=1$, and setting $\tilde B'=d\tilde V'_5$ for 5-form
potential $\tilde V'_5$ (since we may assume that $\tilde H'=0$), we have
\begin{equation}
\hat Q_1 = -\int_{S^{6}}\! d\tilde V'_5 \ .
\label{eq:oneq}
\end{equation}
Defining the 1-form $V$ on the 7-brane's worldvolume by
\begin{equation}
dV = * d{\tilde V}'_5 \ ,
\label{eq:oner}
\end{equation}
we can rewrite (\ref{eq:oneq}) as
\begin{equation}
\hat Q_1 = \int_{S^{6}}\! * dV \ .
\label{eq:ones}
\end{equation}
We conclude that the IIB string may end on an electric charge in a
7-brane. This is just the description of the D-7-brane.


\section{IIA boundaries}

The `brane surgery' method should now be clear. We shall now apply it to 
IIA supergravity, for which the \RR gauge potentials are
\begin{equation}
(C_\mu, A_{\mu\nu\rho})\ ,
\label{eq:twoa}
\end{equation}
i.e. a 1-form $C$ and a 3-form $A$. We might start by considering the CS term
\begin{equation}
F\wedge F\wedge B\ ,
\label{eq:twob}
\end{equation}
where $F$ is the 4-form field strength of $A$. Consideration of this term leads
to the conclusion that (i) a IIA string can end on a 4-brane, and (ii) a
2-brane can end on either a 4-brane or a 5-brane. Since the CS term
(\ref{eq:twob}) is so obviously related to the similar one in D=11 to be
considered below we shall pass over the details. The fact that the IIA string can
also end on either a 2-brane or a 6-brane follows from the fact that the field
strength $F$ has a `modified' Bianchi identity
\begin{equation}
dF = H\wedge K\ ,
\label{eq:twoc}
\end{equation}
where $K=dC$ is the field strength of $C$ (this has a Kaluza-Klein origin in
D=11). We can dualize $A$ to convert this modified Bianchi into a CS term of
the form\footnote{This dualization is inessential to the result, but it allows a
convenient uniformity in the description of the method.}
\begin{equation}
\tilde F \wedge K \wedge B\ ,
\label{eq:twod}
\end{equation} 
where the 6-form $\tilde F$ is, on-shell, the Hodge dual of $F$. This modifies
the $B$ equation to
\begin{equation}
d\star H = -\tilde F \wedge K\ .
\label{eq:twoe}
\end{equation}
We may therefore take the modified charge $\hat Q_1$ to be
\begin{equation}
\hat Q_1 =\int_{S^7} \! [\star H + \tilde F\wedge C]\ .
\label{eq:twof}
\end{equation}
Now, by the identical reasoning used in the IIB case, we first deform the
7-sphere so as to arrive at the formula
\begin{equation}
\hat Q_1 =\int_{S^6}\!  \tilde F\times \int_{S^1}\! C\ .
\label{eq:twog}
\end{equation}
We then identify the first integral as the charge $Q_2$ of a membrane. We then
set $Q_2=1$ and $C=dy$ for some scalar $y$ defined locally on the
worldvolume of the membrane to conclude that the IIA string can end on a
membrane, with the string's charge now being transferred to the
magnetic-type charge 
\begin{equation}
\int_{S^1}\! dy
\label{eq:twoh}
\end{equation}
of a particle on the membrane \cite{asy}. This charge can be non-zero if $y$ is
periodically identified. Clearly, from the KK origin of $C$, we should interpret
$y$ as the coordinate of a hidden 11th dimension. Defining the worldvolume
1-form $V$ by
\begin{equation}
dV= * dy\ ,
\label{eq:twoi}
\end{equation}
we recover \cite{duff,pkt,schm} the usual description of the IIA D-2-brane, in
which the end of the string on the membrane carries the electric charge 
\begin{equation}
\int_{S^1}\! *dV\ .
\label{eq:twoj}
\end{equation}

Returning to (\ref{eq:twoe}) we can alternatively define the modified string
charge to be
\begin{equation}
\hat Q_1 = =\int_{S^7} \! [\star H + K\wedge \tilde A]\ ,
\label{eq:twok}
\end{equation}
where $\tilde A$ is the 5-form potential associated with $\tilde F$, i.e.
$\tilde F= d\tilde A$. Since
\begin{equation}
Q_6=\int_{S^2}\! K
\label{eq:twol}
\end{equation}
is the 6-brane charge, similar reasoning to that above, but now setting $\tilde
A=d\tilde V_4$, leads to the conclusion that a IIA string can also end on a
6-brane and that the string charge is transferred to the 6-brane magnetic charge
\begin{equation}
\int_{S^5} \!d\tilde V_4\ ,
\label{eq:twom}
\end{equation}
which can be rewritten in the expected electric charge form 
\begin{equation}
\int_{S^5} \! * dV
\label{eq:twon}
\end{equation}
by introducing the worldvolume 1-form potential $V$ dual to $\tilde V_4$.

The remaining IIA D-branes are the 0-brane and the 8-brane. The
possibility of a IIA string ending on a 0-brane is {\it not} found by the
`brane surgery' method for the good reason that it is actually forbidden by
charge conservation unless the 0-brane is the endpoint of two or more
strings. Thus, a modification of the method will be needed to deal with this
case. Neither is it it clear how the method  can cope with the IIA 8-brane,
because of the non-generic peculiarities of this case. 

Leaving aside these limitations of the method, there are further consequences to
be deduced from the CS term (\ref{eq:twod}). We have still to consider its
effect on the $\tilde A$ equation of motion. Actually it is easier to return to
the modified Bianchi identity (\ref{eq:twoc}),  which we can rewrite as
\begin{equation}
d(F- K\wedge B)=0\ .
\label{eq:twoo}
\end{equation}
This shows that the homotopy-invariant magnetic 4-brane charge is actually
\begin{equation}
\hat Q_4 = \int_{S^4} [F - K\wedge B]\ .
\label{eq:twop}
\end{equation}
By the now familiar reasoning we deform the 4-sphere and set
$H=0$ to arrive at
\begin{equation}
\hat Q_4 = -\int_{S^2}\! K \times \int_{S^2}\! dV\ .
\label{eq:twoq}
\end{equation}
We recognise the first integral as the charge $Q_6$ of a 6-brane. The second
integral is the magnetic charge associated with a 3-brane within the 6-brane.
The 3-brane is of course the 4-brane's boundary. Thus a 4-brane can end on a
6-brane. This is not unexpected because it follows by T-duality from the
fact that a IIB 3-brane can end on a D-5-brane. 

We could as well have rewritten the modified Bianchi identity (\ref{eq:twoc}) as
\begin{equation}
d(F + H\wedge C)=0\ ,
\label{eq:twor}
\end{equation}
in which case a similar line of reasoning, but setting $K=0$, and so $C=dy$,
leads to the expression
\begin{equation}
\hat Q_4 = \int_{S^3}\! H \times \int_{S^1}\! dy\ .
\label{eq:twos}
\end{equation}
The first integral is the magnetic 5-brane charge $Q_5$, so we deduce that a
4-brane can also end on a (solitonic) 5-brane. The 3-brane boundary in the
5-brane is a magnetic source for the scalar field $y$. The KK origin of $y$
suggests a D=11 interpretation of this possibility. It is surely closely
related to the fact that two D=11 fivebranes can intersect on a 3-brane
\cite{paptown}, since by wrapping one of the 5-branes (but not the other one)
around the 11th dimension we arrive at a D-4-brane intersecting a solitonic
5-brane  in a 3-brane. This is not quite the same as a D-4-brane {\it ending} on
a 5-brane, but the intersection could be viewed as two 4-branes which happen to
end on a common 3-brane boundary in the 5-brane. This illustrates a close
connection between the `brane boundary' rules and the `brane
intersection rules', which will not be discussed here. 


\section{M-brane boundaries}

Finally, we turn to M-theory, or rather D=11 supergravity and its p-brane
solutions. The bosonic fields of D=11 supergravity are the 11-metric and
a 3-form gauge potential $A$ with 4-form field strength $F=dA$. The Bianchi
identity  for $F$ is 
\begin{equation}
dF=0\ ,
\label{eq:threea}
\end{equation}
from which we may immediately conclude that the D=11 fivebrane must be closed.
The same is not true of the D=11 membrane, however, because there is a CS
term in the action of the form
\begin{equation}
{1\over 3} F\wedge F\wedge A
\label{eq:exa}
\end{equation}
which leads to the following field equation\footnote{An additional
singular 5-brane source term was included in \cite{wit} leading to a rather
different interpretation of the significance of the $F\wedge F$ term. We note
that since the 5-brane is actually a completely non-singular solution of the
D=11 field equations \cite{ght} it should not be necessary to include it as a
source.} 
\begin{equation}
d\star F = - F\wedge F\ .
\label{eq:threeb}
\end{equation}
We see that the well-defined membrane charge is actually
\begin{equation}
\hat Q_2 = \int_{S^7}\! [\star F + F\wedge A]\ .
\label{eq:threec}
\end{equation}
Now consider a membrane with a boundary. Contract the 7-sphere to the boundary
and deform it to the product $S^4\times S^3$ so that the entire contibution to
$\hat Q_2$ is given by
\begin{equation}
\hat Q_2 = \int_{S^4}\! F\times \int_{S^3}\! A \ .
\label{eq:threed}
\end{equation}
The first integral is the charge $Q_5$ associated with a fivebrane. Set
$Q_5=1$. We may also set to zero the components of $F$ `parallel' to the
fivebrane, so that $A=dV_2$ in the second integral. We then have
\begin{equation}
\hat Q_2 =  \int_{S^3}\! dV_2\ ,
\label{eq:threee}
\end{equation}
which is the magnetic charge of the string boundary of the membrane in the
fivebrane. 

In fact, the 3-form field-strength $F_3=dV_2$ (or rather ${\cal F}_3 =dV_2-A$
in a general background) is {\it self-dual} but we do not learn this fact from
the `brane surgery' method. As for the IIB 3-brane, where we saw that the
worldvolume 1-forms $V$ and $\tilde V$ are related by Hodge duality of their
2-form field strengths, this information must be gleaned from a different
analysis. The similarity between these constraints on the worldvolume gauge
fields suggests that a deeper understanding of the phenomenon should be
possible. 

In this contribution I have discussed the rules governing `brane boundaries'
in superstring and M-theory and shown that they follow from consideration of
interactions in the corresponding effective supergravity theories. It should be
appreciated that brane boundaries constitute a subset of possible `brane
interactions', which include intersecting branes and branes of varying
topologies. A reasonably complete picture is now emerging of the static aspects
of brane interactions, but little is known at present about the dynamic aspects,
i.e. the analogue of the splitting and joining interaction in string theory.
This problem is presumably bound up with the problem of finding an intrinsic
definition of M-theory, which may well require a substantially new
conceptual framework. Hopefully, the current focus on branes will prove to be of
some help in this daunting task.
\vskip 0.5cm
\noindent
{\it Acknowledgements}: I am grateful to George Papadopoulos for helpful
discussions.

\end{document}